\documentclass[lettersize,twoside,journal]{IEEEtran}
\usepackage{graphicx}
\usepackage{amssymb}
\usepackage{amsmath,cases}
\usepackage{cite}
\usepackage{xcolor}
\usepackage{subfigure}
\usepackage{booktabs}
\usepackage{balance}
\usepackage{acro}
\usepackage{stfloats}
\usepackage{multirow}
\usepackage{mathtools}
\usepackage{etoolbox}

\usepackage[mathscr]{euscript}
\usepackage{colortbl}

\usepackage{microtype}

\usepackage{xcolor}
\usepackage{braket}
\usepackage{bbm}
\usepackage{pifont}

\DeclareMathAlphabet{\mathpzc}{OT1}{pzc}{m}{it}

\DeclareMathAlphabet{\mathitsf}{OML}{cmbr}{m}{it}
\DeclareMathAlphabet{\mathsf}{OT1}{cmbr}{m}{n}

\definecolor{green1}{RGB}{0,51,51}
\definecolor{green2}{RGB}{51,102,102}
\definecolor{green3}{RGB}{125,180,180}
\definecolor{green4}{RGB}{180,218,218}

\definecolor{green0}{RGB}{0,25,25}      
\definecolor{green5}{RGB}{220,238,238}  

\definecolor{red1}{RGB}{220,20,60}  
\definecolor{red2}{RGB}{219,53,47}  

\definecolor{darkgreen}{RGB}{0,128,0}

\newtheorem{definition}{Definition}
\newtheorem{theorem}{Theorem}
\newtheorem{lemma}{Lemma}

\newtheorem{property}{Property}

\AtBeginDocument{}

\newcommand{\B}[1]{\mathbf{#1}}

\newcommand{\R}{\mathbbmss{R}}
\newcommand{\C}{\mathbbmss{C}}

\newcommand{\W}[1]{\M{\mathscr{W}}_{#1}}

\newcommand{\size}[1]{\ensuremath{|#1|}}
\newcommand{\ceil}[1]{\ensuremath{\lceil#1\rceil}}

\newcommand{\Size}[1]{\ensuremath{\left|#1\right|}}

\newcommand{\lra}[1]{\ensuremath{(#1)}}

\newcommand{\lrc}[1]{\ensuremath{\{#1\}}}
\newcommand{\lrd}[1]{\ensuremath{\langle #1\rangle}}

\newcommand{\lrA}[1]{\ensuremath{\left(#1\right)}}
\newcommand{\lrB}[1]{\ensuremath{\left[#1\right]}}
\newcommand{\lrC}[1]{\ensuremath{\left\{#1\right\}}}

\DeclareAcronym{1D}{
short = 1D,
long = one-dimensional
}

\DeclareAcronym{2D}{
short = 2D,
long = two-dimensional
}

\DeclareAcronym{3D}{
short = 3D,
long = three-dimensional
}

\DeclareAcronym{5G}{
short = 5G,
long = fifth-generation
}

\DeclareAcronym{6G}{
short = 6G,
long = sixth-generation
}

\DeclareAcronym{AI}{
short = AI,
long = artificial intelligence
}

\DeclareAcronym{ANN}{
short = ANN,
long = artificial neural network
}

\DeclareAcronym{AR}{
short = AR,
long = augmented reality
}

\DeclareAcronym{AO}{
short = AO,
long = absorptive object
}

\DeclareAcronym{AWGN}{
short = AWGN,
long = additive white Gaussian noise
}

\DeclareAcronym{BB84}{
short = BB84,
long = Bennett and Brassard 1984
}

\DeclareAcronym{BEP}{
short = BEP,
long = bit error probability
}

\DeclareAcronym{BLIP}{
short = BLIP,
long = bootstrapping language-image pretraining
}

\DeclareAcronym{BPSK}{
short = BPSK,
long = binary phase-shift keying
}

\DeclareAcronym{BS}{
short = BS,
long = base station
}

\DeclareAcronym{BSM}{
short = BSM,
long = Bell state measurement
}

\DeclareAcronym{CKA}{
short = CKA,
long = conference key agreement
}

\DeclareAcronym{CNN}{
short = CNN,
long = convolutional neural network
}

\DeclareAcronym{CNOT}{
short = CNOT,
long = controlled-NOT
}

\DeclareAcronym{CPS}{
short = CPS,
long = cyber-physical system
}

\DeclareAcronym{CPTP}{
short = CPTP,
long = completely positive trace-preserving
}

\DeclareAcronym{CPU}{
short = CPU,
long = central processing unit
}

\DeclareAcronym{CSI}{
short = CSI,
long = channel state information
}

\DeclareAcronym{DNN}{
short = DNN,
long = deep neural network
}

\DeclareAcronym{DL}{
short = DL,
long = deep learning
}

\DeclareAcronym{DQL}{
short = DQL,
long = deep Q-learning
}

\DeclareAcronym{DRL}{
short = DRL,
alt = deep RL,
long = deep reinforcement learning
}

\DeclareAcronym{DT}{
short = DT,
long = digital twin
}

\DeclareAcronym{FD}{
short = FD,
long = full-duplex
}

\DeclareAcronym{GAN}{
short = GAN,
long = generative adversarial network
}

\DeclareAcronym{GEO}{
short = GEO,
long = geostationary Earth orbit
}

\DeclareAcronym{GHZ}{
short = GHZ,
long = Greenberger--Horne--Zeilinger
}

\DeclareAcronym{GPS}{
short = GPS,
long = global positioning system
}

\DeclareAcronym{GPT}{
short = GPT,
long = generative pretrained transformer
}

\DeclareAcronym{GPU}{
short = GPU,
long = graphics processing unit 
}

\DeclareAcronym{GUS}{
short = GUS,
long = geometrically uniform symmetry
}

\DeclareAcronym{HQC}{
short = HQC,
long = hybrid quantum-classical
}

\DeclareAcronym{IoT}{
short = IoT,
long = Internet of Things
}

\DeclareAcronym{IIoT}{
short = IIoT,
alt = industrial IoT,
long = industrial Internet of Things
}

\DeclareAcronym{ISAC}{
short = ISAC,
long = integrated sensing and communication
}

\DeclareAcronym{KPI}{
short = KPI,
long = key performance indicator
}

\DeclareAcronym{LEO}{
short = LEO,
long = low-Earth-orbit
}

\DeclareAcronym{LLaMA}{
short = LLaMA,
alt = large language model Meta AI,
long = LLM Meta AI
}

\DeclareAcronym{LLM}{
short = LLM,
long = large language model
}

\DeclareAcronym{LPIPS}{
short = LPIPS,
long = learned perceptual image patch similarity
}

\DeclareAcronym{LOCC}{
short = LOCC,
long = local operations and classical communication
}

\DeclareAcronym{LSTM}{
short = LSTM,
long = long short-term memory
}

\DeclareAcronym{MEO}{
short = MEO,
long = medium-Earth-orbit
}

\DeclareAcronym{MIMO}{
short = MIMO,
long = multiple-input multiple-output
}

\DeclareAcronym{mmWave}{
short = mmWave,
long = millimeter-wave
}

\DeclareAcronym{ML}{
short = ML,
long = machine learning
}

\DeclareAcronym{MLP}{
short = MLP,
long = multi-layer perceptron
}

\DeclareAcronym{MPA}{
short = MPA,
long = mean pixel accuracy
}

\DeclareAcronym{MR}{
short = MR,
long = mixed reality
}

\DeclareAcronym{MSE}{
short = MSE,
long = mean squared error
}

\DeclareAcronym{NISQ}{
short = NISQ,
long = noisy intermediate-scale quantum
}

\DeclareAcronym{NMSE}{
short = NMSE,
alt = normalized MSE,
long = normalized mean squared error
}

\DeclareAcronym{NLP}{
short = NLP,
long = natural language processing
}

\DeclareAcronym{NTN}{
short = NTN,
long = non-terrestrial network
}

\DeclareAcronym{PISQ}{
short = PISQ,
long = perfect intermediate-scale quantum
}

\DeclareAcronym{POVM}{
short = POVM,
long = positive operator-valued measure
}

\DeclareAcronym{PQC}{
short = PQC,
long = parametrized quantum circuit
}

\DeclareAcronym{QAA}{
short = QAA,
long = quantum anonymous authentication
}

\DeclareAcronym{QAB}{
short = QAB,
long = quantum anonymous broadcast
}

\DeclareAcronym{QAC}{
short = QAC,
long = quantum anonymous communication
}

\DeclareAcronym{QACKA}{
short = QA-CKA,
alt = quantum anonymous CKA,
long = quantum anonymous conference key agreement
}

\DeclareAcronym{QACD}{
short = QACD,
long = quantum anonymous collision detection
}

\DeclareAcronym{QAE}{
short = QAE,
long = quantum anonymous entanglement
}

\DeclareAcronym{QAIA}{
short = QAIA,
long = quantum anonymous identity authentication
}

\DeclareAcronym{QAIR}{
short = QAIR,
long = quantum anonymous information retrieval
}

\DeclareAcronym{QAM}{
short = QAM,
long = quadrature amplitude modulation
}

\DeclareAcronym{QAN}{
short = QAN,
long = quantum anonymous network
}

\DeclareAcronym{qAN}{
short = QAN,
long = quantum anonymous notification
}

\DeclareAcronym{QANO}{
short = QANO,
long = quantum anonymous notification
}

\DeclareAcronym{QAP}{
short = QAP,
long = quantum anonymous publication
}

\DeclareAcronym{QAR}{
short = QAR,
long = quantum anonymous ranking
}

\DeclareAcronym{QAS}{
short = QAS,
long = quantum anonymous sensing
}

\DeclareAcronym{QAT}{
short = QAT,
long = quantum anonymous teleportation
}

\DeclareAcronym{QAV}{
short = QAV,
long = quantum anonymous voting
}

\DeclareAcronym{QCRB}{
short = QCRB,
long = quantum Cram\'er--Rao bound
}

\DeclareAcronym{QCS}{
short = QCS,
long = quantum clock synchronization
}

\DeclareAcronym{QEC}{
short = QEC,
long = quantum error correction
}

\DeclareAcronym{QED}{
short = QED,
long = quantum entanglement distribution
}

\DeclareAcronym{QFI}{
short = QFI,
long = quantum Fisher information
}

\DeclareAcronym{QFIM}{
short = QFIM,
alt = QFI matrix,
long = quantum Fisher information matrix
}

\DeclareAcronym{QFT}{
short = QFT,
long = quantum Fourier transform
}

\DeclareAcronym{QIoT}{
short = QIoT,
alt = quantum IoT,
long = quantum Internet of Things
}

\DeclareAcronym{QKD}{
short = QKD,
long = quantum key distribution
}

\DeclareAcronym{QML}{
short = QML,
alt = quantum ML,
long = quantum machine learning
}

\DeclareAcronym{QNN}{
short = QNN,
long = quantum neural network
}

\DeclareAcronym{QOC}{
short = QOC,
long = quantum optimal control
}

\DeclareAcronym{QPU}{
short = QPU,
long = quantum processing unit
}

\DeclareAcronym{QSC}{
short = QSC,
alt = quantum SC,
long = quantum semantic communication
}

\DeclareAcronym{QSN}{
short = QSN,
long = quantum sensing network
}

\DeclareAcronym{ReLU}{
short = ReLU,
long = rectified linear unit
}

\DeclareAcronym{ResNet}{
short = ResNet,
long = residual network
}

\DeclareAcronym{RGB}{
short = RGB,
long = red-green-blue
}

\DeclareAcronym{RL}{
short = RL,
long = reinforcement learning
}

\DeclareAcronym{RNN}{
short = RNN,
long = recurrent neural network
}

\DeclareAcronym{RSA}{
short = RSA,
long = Rivest--Shamir--Adleman
}

\DeclareAcronym{SAM}{
short = SAM,
long = segment anything model
}

\DeclareAcronym{SC}{
short = SC,
long = semantic communication
}

\DeclareAcronym{SEP}{
short = SEP,
long = symbol error probability
}

\DeclareAcronym{SNR}{
short = SNR,
long = signal-to-noise ratio
}

\DeclareAcronym{SRM}{
short = SRM,
long = square-root measurement
}

\DeclareAcronym{SQL}{
short = SQL,
long = standard quantum limit
}

\DeclareAcronym{SVM}{
short = SVM,
long = support vector machine
}

\DeclareAcronym{SwinT}{
short = SwinT,
long = swin (shifted window) transformer
}

\DeclareAcronym{THz}{
short = THz,
long = terahertz
}

\DeclareAcronym{TN}{
short = TN,
long = terrestrial network
}

\DeclareAcronym{UAV}{
short = UAV,
long = unmanned aerial vehicle
}

\DeclareAcronym{U-Net}{
short = U-Net,
long = U-shaped network
}

\DeclareAcronym{VAE}{
short = VAE,
long = variational autoencoder
}

\DeclareAcronym{V2I}{
short = V2I,
long = vehicle-to-infrastructure
}

\DeclareAcronym{V2X}{
short = V2X,
long = vehicle-to-everything
}

\DeclareAcronym{VFM}{
short = VFM,
long = vision foundation model
}

\DeclareAcronym{ViT}{
short = ViT,
long = vision transformer
}

\DeclareAcronym{VQA}{
short = VQA,
long = variational quantum algorithm
}

\DeclareAcronym{VQC}{
short = VQC,
long = variational quantum circuit
}

\DeclareAcronym{VQS}{
short = VQS,
long = variational quantum sensing
}

\DeclareAcronym{VR}{
short = VR,
long = virtual reality
}

\DeclareAcronym{YOLO}{
short = YOLO,
long = you only look once
}

\usepackage{tikz}
\usetikzlibrary{calc,positioning}

\tikzset{
	timeline axis/.style={line width=1pt},
	timeline tick/.style={line width=1pt},
	timeline guide/.style={densely dotted, line width=0.9pt},
}

\newcounter{tlflip}

\newcommand{\entry}[4][]{%
	\pgfmathsetmacro{\stage}{#2}%
	\pgfmathsetmacro{\x}{#3-\tlstart}%
	\pgfmathsetmacro{\y}{0.9*\stage}%
	
	\draw[timeline tick] (\x,0.2) -- (\x,-0.2);
	
	\def\side{#1}%
	\def\sideup{up}%
	\def\sidedown{down}%
	\def\sideupleft{upleft}%
	\def\sideupright{upright}%
	\def\sidedownleft{downleft}%
	\def\sidedownright{downright}%
	
	\ifx\side\empty
	\ifodd\value{tlflip}
	\def\side{up}%
	\else
	\def\side{down}%
	\fi
	\fi
	
	\ifx\side\sideup
	\node[above,align=center] at (\x,0.35) {#3};
	\draw[timeline guide] (\x,0.90) -- (\x,\y);
	\node[above,align=center] at (\x,\y) {#4};
	
	\else\ifx\side\sidedown
	\node[below,align=center] at (\x,-0.35) {#3};
	\draw[timeline guide] (\x,-0.90) -- (\x,-\y);
	\node[below,align=center] at (\x,-\y) {#4};
	
	\else\ifx\side\sideupleft
	\node[above,align=center] at (\x,0.35) {#3};
	\pgfmathsetmacro{\xt}{\x-3.0} 
	\draw[timeline guide] (\x,0.9) -- (\xt+0.2,\y-0.2); 
	\node[above,align=center] at (\xt,\y) {#4};
	
	\else\ifx\side\sideupright
	\node[above,align=center] at (\x,0.35) {#3};
	\pgfmathsetmacro{\xt}{\x+3.0}
	\draw[timeline guide] (\x,0.9) -- (\xt-0.2,\y-0.2);
	\node[above,align=center] at (\xt,\y) {#4};
	
	\else\ifx\side\sidedownleft
	\node[below,align=center] at (\x,-0.35) {#3};
	\pgfmathsetmacro{\xt}{\x-3.0}
	\draw[timeline guide] (\x,-0.9) -- (\xt+0.2,-\y+0.2);
	\node[below,align=center] at (\xt,-\y) {#4};
	
	\else\ifx\side\sidedownright
	\node[below,align=center] at (\x,-0.35) {#3};
	\pgfmathsetmacro{\xt}{\x+3.0}
	\draw[timeline guide] (\x,-0.9) -- (\xt-0.2,-\y+0.2);
	\node[below,align=center] at (\xt,-\y) {#4};
	
	\else
	\node[above,align=center] at (\x,0.35) {#3};
	\draw[timeline guide] (\x,0.90) -- (\x,\y);
	\node[above,align=center] at (\x,\y) {#4};
	\fi\fi\fi\fi\fi\fi
	
	\addtocounter{tlflip}{1}%
}

\tikzset{
	vtaxis/.style={line width=1.0pt},
	vtmark/.style={line width=1.0pt},
	vtlink/.style={line width=0.7pt},
}

\newcounter{vtLine}        
\newcounter{vtTotalLines}  

\def\vtRowSep{0.45}   
\def\vtStartY{0}       

\def\vtAllCount{}
\def\vtAllDraw{}
\def\vtlastyear{*}

\makeatletter
\newcount\vt@linecount

\def\vtCountLines#1{%
	\vt@linecount=0\relax
	\expandafter\vtCountLinesAux#1\\\vt@nil
}
\def\vtCountLinesAux#1\\#2\vt@nil{%
	\advance\vt@linecount by 1\relax
	\def\vt@next{#2}%
	\ifx\vt@next\empty
	\else
	\expandafter\vtCountLinesAux#2\vt@nil
	\fi
}

\def\vtDrawLines#1{%
	\expandafter\vtDrawLinesAux#1\\\vt@nil
}
\def\vtDrawLinesAux#1\\#2\vt@nil{%
	\stepcounter{vtLine}%
	\pgfmathsetmacro{\idx}{\value{vtLine}-1}%
	\pgfmathsetmacro{\y}{\vtStartY + \idx*\vtRowSep}%
\node[anchor=west, align=left] at (\vtTextX,\y) {#1};

	\def\vt@next{#2}%
	\ifx\vt@next\empty
	\else
	\expandafter\vtDrawLinesAux#2\vt@nil
	\fi
}

\newcommand{\vtDoEntry}[2]{%
	\edef\thisyear{#1}%
	\pgfmathsetmacro{\ybase}{\vtyearscale*(#1-\vtstartyear)}%
	
	\ifx\thisyear\vtlastyear
	\else
	\node[anchor=east] at (\vtYearX,\ybase) {#1};
	\draw[vtmark] (\vtAxisX,\ybase) -- (\vtAxisX+0.01,\ybase);
	\edef\vtlastyear{#1}%
	\fi
	
	\pgfmathsetmacro{\idxfirst}{\value{vtLine}}%
	\pgfmathsetmacro{\yfirst}{\vtStartY + \idxfirst*\vtRowSep}%
	\draw[vtlink] (\vtAxisX+0.01,\ybase) -- (\vtDiagX,\yfirst) -- (\vtTextX,\yfirst);
	
	\vtDrawLines{#2}%
}

\newcommand{\vtCountEntry}[2]{%
	\vtCountLines{#2}%
	\addtocounter{vtTotalLines}{\the\vt@linecount}%
}

\newcommand{\ventry}[2]{%
	\g@addto@macro\vtAllCount{\vtCountEntry{#1}{#2}}%
	\g@addto@macro\vtAllDraw{\vtDoEntry{#1}{#2}}%
}
\makeatother

\usepackage{makecell}
\usepackage{comment}
\usepackage{threeparttable}
\usepackage[hidelinks]{hyperref}

\usepackage{algorithm}
\usepackage{algorithmic}

\usepackage{booktabs}
\usepackage{fixmath}
\usepackage{bm}
\usepackage{amsbsy}
\usepackage{color}
\usepackage{verbatim}
\usepackage{array}
\usepackage{tikz}
\usetikzlibrary{arrows.meta, positioning, calc}

\usetikzlibrary{calc}
\usetikzlibrary{decorations.pathmorphing}
\def\A{\mathcal{A}}
\def\B{\mathcal{B}}
\def\C{\mathcal{C}}
\def\E{\mathcal{E}}

\def\G{\mathcal{G}}

\def\I{\mathcal{I}}

\def\P{\mathcal{P}}

\def\S{\mathcal{S}}
\def\T{\mathcal{T}}
\def\U{\mathcal{U}}
\def\V{\mathcal{V}}
\def\W{\mathcal{W}}

\allowdisplaybreaks

\def\OPT{\mathsf{OPT}}
   
\DeclareAcronym{BTCP}{
short = B-TCP,
long = budgeted tour cover problem
}

\DeclareAcronym{BTSP}{
short = BTSP,
alt = biased TSP,
long = biased traveling salesman problem
}

\DeclareAcronym{BSCP}{
short = B-SCP,
alt = budgeted SCP,
long = budgeted sweep coverage problem
}

\DeclareAcronym{CFA}{
short = CFA,
long = constant-factor approximation
}

\DeclareAcronym{CWBSCP}{
short = CWB-SCP,
long = cyclic weighted budgeted sweep coverage problem
}

\DeclareAcronym{FPT}{
short = FPT,
long = fixed-parameter tractable
}

\DeclareAcronym{HWBSCP}{
short = HWB-SCP,
long = heterogeneous weighted budgeted sweep coverage problem
}

\DeclareAcronym{LP}{
short = LP,
long = linear programming
}

\DeclareAcronym{MCP}{
short = MCP,
long = maximum coverage problem
}

\DeclareAcronym{MEP}{
short = MEP,
long = minimum-excess path
}

\DeclareAcronym{GBMCP}{
short = GB-MCP,
alt = group-budget MCP,
long = group-budget maximum coverage problem
}

\DeclareAcronym{MCG}{
short = MCG,
alt = MCP with group budget constraints,
long = maximum coverage problem with group budget constraints
}

\DeclareAcronym{MSSCP}{
short = MS-SCP,
long = min-sensor sweep coverage problem
}
\DeclareAcronym{MMSPP}{
short = MM-SPP,
long = min-max sweep period problem
}
\DeclareAcronym{MOP}{
short = $M$-OP,
long = $M$-orienteering problem
}

\DeclareAcronym{OP}{
short = OP,
long = orienteering problem
}

\DeclareAcronym{MST}{
short = MST,
long = minimum-cost spanning tree
}

\DeclareAcronym{PCB}{
short = PCB,
long = printed circuit board
}
\DeclareAcronym{PoI}{ 
  short     = PoI, 
  long      = point of interest,
  long-plural-form = points of interest,
  short-plural = s
}
\DeclareAcronym{PSCP}{
short = P-SCP,
long = partial sweep coverage problem
}
\DeclareAcronym{PCSCP}{
short = PC-SCP,
long = prize-collecting sweep coverage problem
}

\DeclareAcronym{SCP}{
short = SCP,
long = the sweep coverage problem
}
\DeclareAcronym{SGPP}{
short = SGPP,
long = subgroup path planning problem
}
\DeclareAcronym{SPP}{
short = SPP,
long = the subpath planning problem
}

\DeclareAcronym{TSP}{
short = TSP,
long = traveling salesman problem
}
\DeclareAcronym{TSPP}{
short = TSPP,
long = traveling salesman path problem
}
\DeclareAcronym{TSPPK}{
short = $k$-TSPP,
long = $k$-person traveling salesman path problem
}

\DeclareAcronym{WBSCP}{
short = WB-SCP,
long = weighted budgeted sweep coverage problem
}
\DeclareAcronym{WPSCP}{
short = WP-SCP,
long = weighted partial sweep coverage problem
}

\DeclareAcronym{XP}{
short = XP,
long = slice-wise polynomial-time
}

\hypersetup{hypertexnames=false}

\makeatletter
\patchcmd{\float@makebox}{\@parboxrestore}{\@parboxrestore\bgroup\color{black}}{}{}
\patchcmd{\float@makebox}{\vskip\z@}{\vskip\z@\egroup}{}{}
\makeatother

\AtBeginEnvironment{thebibliography}{%
  \let\BSCoriginalbibitem\bibitem
  \renewcommand{\bibitem}[2][]{%
    \ifstrequal{#2}{BCKLMM:07:SICOMP}{\color{blue}}{\color{black}}
    \ifstrempty{#1}{\BSCoriginalbibitem{#2}}{\BSCoriginalbibitem[#1]{#2}}%
  }%
}

\begin{document}

\title{Constant-Factor Approximation Algorithms for Weighted Budgeted Multisensor Sweep Coverage}
\author{
Jingyang~Zhao,
Jason William Setiawan,
Uman Khalid,
Chaoning~Zhang,~\IEEEmembership{Senior~Member,~IEEE},
Een-Kee~Hong,~\IEEEmembership{Senior~Member,~IEEE},
and
Hyundong~Shin,~\IEEEmembership{Fellow,~IEEE}
 \thanks{
J.~Zhao, J.~Setiawan, U.~Khalid, E.-K.~Hong, and H.~Shin (corresponding author) are with the Department of Electronics and Information Convergence Engineering, Kyung Hee University, 1732 Deogyeong-daero, Giheung-gu, Yongin-si, Gyeonggi-do 17104, Republic of Korea (e-mail: hshin@khu.ac.kr).
 }
\thanks{
C.~Zhang is with the School of Computer Science and Engineering, University of Electronic Science and Technology of China, Chengdu, China (e-mail: chaoningzhang@uestc.edu.cn).
}
 }
\markboth{}{ 
 Zhao \textit{\MakeLowercase{et al.}}:
Constant-Factor Approximation Algorithms for Weighted Budgeted Multisensor Sweep Coverage 
 }

\maketitle

\begin{abstract}

\textls[-1]{Sweep coverage requires each \ac{PoI} to be revisited by some mobile sensor at least once during every sweep period, and it underlies persistent patrolling, environmental monitoring, and infrastructure inspection. Given a set of $N$ \acp{PoI}, each associated with a nonnegative weight representing its importance, the \ac{WBSCP} seeks routing trajectories for a budget of $M\leq N$ identical mobile sensors with a common maximum speed to maximize the total weight of the sweep-covered \acp{PoI}. For the unweighted special case in which every \ac{PoI} has unit weight, the first constant-factor approximation algorithm achieves a ratio of $0.0116-O\lrA{\varepsilon}$ for any constant $\varepsilon>0$, and whether the \ac{WBSCP} itself admits such an algorithm has remained open. In this paper, we resolve this open question by proposing two constant-factor approximation algorithms for the \ac{WBSCP}.
The first converts any bifactor $\lrA{\alpha,\beta}$-approximation algorithm into a $\beta/\lrA{2\alpha}$-approximation algorithm, yielding a ratio of $1/16=0.0625$.
The second builds on a greedy maximum-coverage framework and improves the ratio to $1-e^{-1/2}-\varepsilon$ for any fixed $\varepsilon>0$.
Both algorithms also improve the best-known ratio for the unweighted case, and the second extends to heterogeneous sensors with different speeds while preserving the same guarantee.}

\end{abstract}

 \begin{IEEEkeywords}
 Approximation algorithms,
 combinatorial optimization,
 multisensor systems,
 routing, 
 sweep coverage.
  \end{IEEEkeywords}



\acresetall		


\section{Introduction}\label{sec:introduction}

\IEEEPARstart{M}{obile} robotics and wireless sensor networks have long studied the target coverage problem because of its broad applications in intelligent surveillance, smart-city patrolling, environmental monitoring, infrastructure inspection, and disaster response~\cite{LCHHLL:11:IEEE_J_MC,GFWC:18:IEEE_J_NET,LD:21:IEEE_J_IOT,GFWC:22:IEEE_J_MC,KSSD:24:IEEE_J_NET,Xu_etal:24:IEEE_J_NET,Guo_etal:22:IEEE_J_MC,Shen_etal:23:IEEE_J_IOT,RHZD:21:J_Global_Optim,Yang_etal:22:IEEE_J_NET,AB:23:IEEE_J_CYB,WZMZ:23:IEEE_J_CYB,HVN:20:IEEE_J_CYB,PKLM:20:IEEE_J_CYB,ZKPZZS:26:IEEE_O_CSTO}.
Traditional deployment paradigms rely on a fixed sensing infrastructure to provide continuous monitoring for a given set of \acp{PoI}~\cite{YK:22:IEEE_J_CYB}. 
Although static sensors provide continuous local monitoring, it scales poorly in large-scale or sparsely distributed environments~\cite{XC:22:IEEE_J_ITS,LXSW:20:IEEE_J_IOT,CCLX:23:IEEE_J_SMC}, incurring substantial deployment, maintenance, and battery replacement costs~\cite{GCPWC:20:IEEE_J_MC,WXWHS:19:IEEE_J_IOT}.

To overcome these limitations, the concept of \emph{sweep coverage} has been introduced as a dynamic alternative. Under this paradigm, a fleet of identical mobile sensors follows prescribed trajectories to visit the \acp{PoI} periodically. Each \ac{PoI} is associated with a required sweep period, and is said to be \emph{sweep-covered} if it is visited at least once during every sliding time window of that length. Throughout, $N$ denotes the number of \acp{PoI} and $M$ the number of available sensors. 
This sweep-coverage paradigm arises in numerous real-world scenarios, including autonomous ground robots patrolling office complexes, unmanned aerial vehicles collecting multispectral imagery over agricultural fields, and autonomous surface vessels monitoring water quality in urban reservoirs.

\subsection{Background and Motivation}

\textls[-4]{Early theoretical studies on sweep coverage have primarily focused on minimizing the number of sensors required to sweep-cover all \acp{PoI}~\cite{LCHHLL:11:IEEE_J_MC}.
In many practical deployments, however, the number of available sensors is limited and often insufficient to sweep-cover every \ac{PoI}. Consequently, increasing attention has been devoted to maximizing the coverage achieved by a limited number of sensors.
This setting gives rise to the budgeted \ac{SCP} (\acs{BSCP})~\cite{HZLSYG:18:IEEE_Access,LZD:24:Optim_Lett,WSZ:25:IEEE_J_NET}. 
In the standard \ac{BSCP}, all \acp{PoI} share a common sweep period, which bounds the distance each sensor can travel within one period.
The objective is to compute a feasible schedule that maximizes the number of sweep-covered \acp{PoI} using a given fleet of $M$ sensors. When each \ac{PoI} is additionally assigned a nonnegative weight representing its priority, importance, or monitoring reward, the problem generalizes to the weighted \ac{BSCP} (\acs{WBSCP}), whose objective is to maximize the total weight of the sweep-covered \acp{PoI}. Unlike the unweighted case, the weighted objective prevents the direct application of existing cardinality-based approximation techniques.}

A $\rho$-approximation algorithm is a polynomial-time algorithm that always returns a solution whose objective value is within (respectively, at least) a factor $\rho$ of the optimum for minimization (respectively, maximization) problems, and $\rho$ is called the \emph{approximation ratio}.
The \ac{BSCP} was first investigated in~\cite{HZLSYG:18:IEEE_Access}, which proposes a $\lra{1-e^{-1}}$-approximation algorithm under the assumption that each sensor selects its trajectory from a polynomial-sized candidate set. 
Under this assumption, the routing aspect of the problem is essentially removed, and the problem reduces to the classical \ac{MCP}~\cite{CA:04:APPROX,CM:18:ALG}, allowing the standard greedy algorithm and \ac{LP} rounding techniques to be applied.

Subsequently, two heuristic algorithms for the \ac{WBSCP} are given in~\cite{NLD:20:COCOA}.
The heterogeneous \ac{WBSCP} on line metrics, where all \acp{PoI} lie on a line, is investigated in~\cite{LS:21:Sensors}, where substantially stronger results are obtained by exploiting the special geometric structure. Specifically, this work proposes a $\lra{1-e^{-1}}/2$-approximation algorithm for arbitrary sensor speeds, a $1/2$-approximation algorithm when the number of distinct sensor speeds is constant, and an exact algorithm when all sensors have the same speed.
For general metric graphs, however, substantially less is known.
For the \ac{BSCP} on general metrics, the best-known result for several years was a bifactor $\lrA{4,1/2}$-approximation algorithm~\cite{LZD:24:Optim_Lett}, which computes a solution using at most $4M$ sensors while guaranteeing at least half of the optimal coverage obtainable by $M$ sensors. This bifactor approximation framework also extends to the \ac{WBSCP} and remains the strongest known theoretical guarantee for it.

Whether the \ac{BSCP} admits a constant-factor approximation algorithm is raised as an open question in~\cite{LZD:24:Optim_Lett}. 
Recently, this question has been answered in~\cite{WSZ:25:IEEE_J_NET} by introducing the \acl{MOP} and establishing a reduction showing that any $\rho$-approximation algorithm for the \acl{MOP} yields a $\rho/3$-approximation algorithm for the \ac{BSCP}. 
By further developing a constant-factor approximation algorithm for the \acl{MOP}, this work obtains the first constant-factor approximation algorithm for the unweighted \ac{BSCP}, achieving an approximation ratio of $0.0116-O\lrA{\varepsilon}$ for any constant $\varepsilon>0$.
Although this breakthrough establishes the existence of a constant-factor approximation algorithm, the resulting approximation ratio remains rather small.
Moreover, this framework relies on cardinality-based arguments and therefore does not extend to the weighted setting. As pointed out in~\cite{WSZ:25:IEEE_J_NET}, whether the \ac{WBSCP} admits a constant-factor approximation algorithm remains an open problem.
Beyond the weighted setting, another natural direction is to consider the \ac{WBSCP} with heterogeneous sensors having different travel speeds, which is also posed as an open problem in~\cite{WSZ:25:IEEE_J_NET}. Existing approximation algorithms on general metrics assume identical sensors moving at a common speed, and extending them to heterogeneous travel speeds remains challenging.

\subsection{Related Work} 

Sweep coverage has evolved into several theoretical variants with different optimization objectives.
The minimum-sensor \ac{SCP}, introduced in~\cite{LCHHLL:11:IEEE_J_MC}, aims to find a feasible sweep-coverage solution using the minimum number of sensors while satisfying the individual sweep-period requirements of all \acp{PoI}. On general graphs, achieving an approximation ratio better than $2$ is NP-hard~\cite{LCHHLL:11:IEEE_J_MC}.
Subsequently, a $3$-approximation algorithm based on minimum-cost spanning forests to cluster nearby targets is given in~\cite{GM:15:Inf_Process_Lett}. 
When \acp{PoI} have heterogeneous sweep-period requirements, the same algorithm achieves an approximation ratio of $O\lrA{\rho}$, where $\rho$ denotes the ratio between the maximum and minimum sweep periods.
A closely related problem minimizes the maximum sweep period while sweep-covering all \acp{PoI} using $M$ sensors. A $4$-approximation algorithm for this problem is known~\cite{GFWC:22:IEEE_J_MC}.
Moreover, a $\lra{5-2/\lra{N-M+1}}$-approximation algorithm is available for the setting in which each sensor independently traverses a cyclic trajectory~\cite{GFWC:18:IEEE_J_NET}.

The prize-collecting \ac{SCP} is introduced in~\cite{LZ:21:AAIM}. In this problem, each uncovered \ac{PoI} incurs a penalty, and the objective is to minimize the sum of the deployment cost of sensors and the penalties of uncovered \acp{PoI}. An $8$-approximation algorithm is given in~\cite{LZD:24:Optim_Lett}, and the approximation ratio has recently been improved to $5$~\cite{LTZ:25:Math_Program}. Moreover, a $5$-approximation algorithm is available for the case where sensors are dispatched from a constant number of depots~\cite{LZ:21:AAIM}.
Finally, the partial \ac{SCP} (or its weighted version) seeks a minimum number of sensors to sweep-cover at least a given number (or total weight) of \acp{PoI}. An $8$-approximation algorithm for both problems is known~\cite{LZD:24:Optim_Lett}.

{
\color{black}
\subsection{Contributions}

In this paper, we substantially improve the state-of-the-art for both the \ac{BSCP} and the \ac{WBSCP}. Our main contributions are summarized as follows.

\subsubsection{Simple Constant-Factor Approximation for the \ac{WBSCP}}

\textls[-2]{We first propose a simple constant-factor approximation algorithm for the \ac{WBSCP}, achieving an approximation ratio of $1/16$. This result already resolves an open problem raised in~\cite{WSZ:25:IEEE_J_NET}. Since the \ac{BSCP} is a special case of the \ac{WBSCP}, this algorithm also improves the best-known approximation ratio for the \ac{BSCP} from $0.0116-O\lrA{\varepsilon}$ to $1/16=0.0625$.
The algorithm is built upon a generic transformation applicable to any bifactor $\lrA{\alpha,\beta}$-approximation algorithm for the \ac{WBSCP}. 
The key ingredient is a simple yet powerful structural observation showing that the collaborative sweep routes produced by a bifactor approximation algorithm can be converted into independent periodic routes while increasing the number of required sensors by at most a factor of two. This transformation turns any $\lrA{\alpha,\beta}$-approximation algorithm for the \ac{WBSCP} into a $\beta/\lra{2\alpha}$-approximation algorithm. In particular, by using the $\lrA{4,1/2}$-approximation algorithm~\cite{LZD:24:Optim_Lett}, we obtain a $1/16$-approximation algorithm.}

\subsubsection{Improved Approximation via a Greedy Framework}

We develop a greedy approximation framework that achieves an approximation ratio of $\lra{1-e^{-1/2}-\varepsilon}$ for any fixed $\varepsilon>0$, improving the approximation ratio for both the \ac{WBSCP} and the \ac{BSCP}.
Specifically, short-window averaging relates the weight covered by an arbitrary feasible schedule to short paths.
Using a bifactor approximation for weighted minimum-excess paths~\cite{FS:14:STOC}, we construct budget-feasible tours that retain nearly all the weight of these paths.
The residual-weight greedy framework then selects at most $M$ tours, one per sensor, and attains the stated guarantee in polynomial time for fixed $\varepsilon$.

\subsubsection{Extension to Heterogeneous Sensor Speeds}

We further extend our framework to the heterogeneous \ac{WBSCP} (\acs{HWBSCP}), which considers fleets of sensors with heterogeneous travel speeds, thereby addressing another open problem raised in~\cite{LS:21:Sensors}.
Although heterogeneous distance budgets invalidate the simple greedy strategy used in the homogeneous setting, we consider a group-coverage formulation based on the \aca{GBMCP} (\ac{GBMCP}), where each sensor defines a separate group.
The same short-window bound and route oracle satisfy the conditions of a group-coverage lemma, which selects at most one feasible tour for each sensor.
This yields a $\lra{1-e^{-1/2}-\varepsilon}$-approximation algorithm for any fixed $\varepsilon>0$.
Moreover, on line and cycle metrics, by enumerating a group of feasible coverage sets for each sensor and applying coverage \ac{LP} rounding, we obtain a $\lra{1-e^{-1/2}}$-approximation algorithm.
On line metrics, this ratio improves the previous $\lra{1-e^{-1}}/2$ guarantee in~\cite{LS:21:Sensors}, while extending the same ratio to cycle metrics.

A summary of previous and our results is provided in Table~\ref{res}, where $\alpha=4$ and $\beta=1/2$ are the factors in the current best-known bifactor $\lrA{4,1/2}$-approximation algorithm for the \ac{WBSCP}~\cite{LZD:24:Optim_Lett}.

\begin{table}[t]
\centering
\caption{Approximation algorithms for the \ac{BSCP} and its variants}
\begin{threeparttable}
\setlength{\tabcolsep}{3.5pt}
\begin{tabular}{llll}
\toprule
Problem & Approximation & Running Time & Reference\\
\midrule
\midrule
\multirow{5}*{\ac{BSCP}}
& $\lra{4, 1/2}$ & Polynomial & \cite{LZD:24:Optim_Lett} \\
\cmidrule{2-4}
& $0.0116-O\lrA{\varepsilon}$ & Polynomial & \cite{WSZ:25:IEEE_J_NET} \\
\cmidrule{2-4}
& $\beta/(2\alpha)$ $(=1/16)$ & Polynomial & Theorem~\ref{thm1} \\
\cmidrule{2-4}
& $1-e^{-1/2}-\varepsilon$ & Polynomial & Theorem~\ref{thm3} \\
\midrule
\multirow{4}*{\ac{WBSCP}}
& $\lra{4, 1/2}$ & Polynomial & \cite{LZD:24:Optim_Lett} \\
\cmidrule{2-4}
& $\beta/(2\alpha)$ $(=1/16)$ & Polynomial & Theorem~\ref{thm1} \\
\cmidrule{2-4}
& $1-e^{-1/2}-\varepsilon$ & Polynomial & Theorem~\ref{thm3} \\
\midrule
\multirow{4}*{\ac{HWBSCP}} & $\lra{1-e^{-1}}/2$
& Polynomial\,$^{\text{a)}}$ & \cite{LS:21:Sensors} \\
\cmidrule{2-4}
& $1-e^{-1/2}-\varepsilon$ & Polynomial & Theorem~\ref{thm5} \\
& $1-e^{-1/2}$ & Polynomial\,$^{\text{a),\ b)}}$ & Theorem~\ref{cor:plus}\\
\bottomrule
\end{tabular}
\begin{tablenotes}
\item[a)] Line metrics; $^{\text{b)}}$\,Cycle metrics.
\end{tablenotes}
\end{threeparttable}
\label{res}
\end{table}
}

\subsection{Organization}

The remainder of the paper is organized as follows.
Section~\ref{sec:2} introduces the \ac{SCP} preliminaries, including the formulation of the \ac{WBSCP}.
Section~\ref{sec:3} presents our $1/16$-approximation algorithm for the \ac{WBSCP}, including its trajectory extraction, sequential shortcutting, and cycle decomposition techniques.
Section~\ref{sec:4} provides our improved $\lra{1-e^{-1/2}-\varepsilon}$-approximation algorithm for the \ac{WBSCP}, including short-window averaging, the minimum-excess path subroutine, and the associated greedy framework and approximation analysis.
Section~\ref{sec:5} extends our framework to the \ac{HWBSCP} for heterogeneous sensors, including a group-coverage formulation that yields a $\lra{1-e^{-1/2}-\varepsilon}$-approximation algorithm in general, alongside a $\lra{1-e^{-1/2}}$-approximation algorithm specific to line and cycle metrics.
Section~\ref{sec:group_coverage} proves the general group-coverage lemma.
Section~\ref{sec:7} concludes the paper.

\section{Preliminaries} \label{sec:2}

\textls[-1]{An instance of the \ac{WBSCP} is formulated on a complete metric graph $\G = \lrA{\V, \E}$, where the vertex set $\V = \lrC{v_1, \ldots, v_N}$ represents the set of $N$ \acp{PoI} distributed across a monitoring space, and $\E$ represents the set of edges connecting them.
The graph is associated with a metric edge cost function $c:\E\rightarrow\R_{\ge0}$ satisfying $c\lrA{x,x}=0$, $c\lrA{x,y}=c\lrA{y,x}$, and $c\lrA{x,z}\le c\lrA{x,y}+c\lrA{y,z}$ (the triangle inequality) for all $x,y,z\in\V$.
Furthermore, to capture the heterogeneous significance of different \acp{PoI}, each vertex $v\in\V$ is assigned a nonnegative weight $w\lrA{v}$.}

A \emph{walk} $W=\lrd{v_1,v_2,\ldots,v_\ell}$ is defined as a sequence of (not necessarily distinct) vertices.
Denote its edge multiset, in which repeated edges appear with multiplicity, and its total cost, respectively, by
\begin{align}
\E\lrA{W}
&\coloneqq\lrC{v_1v_2,v_2v_3,\ldots,v_{\ell-1}v_\ell}, \\
c\lrA{W}
&\coloneqq\sum_{e\in \E\lrA{W}}c\lrA{e},
\end{align}
and the set of vertices it visits by $\V\lrA{W}$.
The walk $W$ is called a \emph{tour} if $v_1=v_\ell$, and it is called a \emph{path} if all its vertices are distinct.
A tour is called a \emph{cycle} if it can be written as $\lrd{v_1,v_2,\ldots,v_{\ell-1},v_1}$, where the vertices $v_1,\ldots,v_{\ell-1}$ are pairwise distinct.
For instance, $\lrd{v_1,v_2,v_1,v_3,v_1}$ is a tour but not a cycle, whereas $\lrd{v_1,v_2,v_3}$ and $\lrd{v_1,v_2,v_3,v_1}$ are a path and a cycle, respectively.
Additionally, two walks $W$ and $W'$ are said to be vertex-disjoint if $\V\lrA{W}\cap\V\lrA{W'}=\emptyset$.

\textls[-1]{For any vertex subset $\S\subseteq\V$, we define its weight as $w\lrA{\S}\coloneqq\sum_{v\in\S}w\lrA{v}$.
For any set $\W$ of walks, the covered vertex set is denoted by $\V\lrA{\W}\coloneqq\bigcup_{W\in\W}\V\lrA{W}$, yielding a total collected weight of $w\lrA{\W}\coloneqq w\lrA{\V\lrA{\W}}$. 
We also define $w\lrA{W}\coloneqq w\lrA{\V\lrA{W}}$ for any walk $W$ and $\lrB{\ell}\coloneqq\lrC{1,2,\ldots,\ell}$ for any positive integer $\ell$.
We allow empty paths and tours, both with empty vertex sets and zero cost and weight. Singleton paths, tours, and cycles have zero cost.
Given a nonempty path $P=\lrd{v_1,\ldots,v_\ell}$, a cycle $C=\lrd{v_1,\ldots,v_\ell,v_1}$ can be obtained by adding the edge $v_\ell v_1$. This operation is called \emph{completing} the path $P$. By the triangle inequality, $c\lrA{C}\leq 2c\lrA{P}$. 
Moreover, given a walk $W$, a new walk $W'$ can be obtained by skipping any repeated vertex. This operation is called \emph{shortcutting}. Again, by the triangle inequality, $c\lrA{W'}\leq c\lrA{W}$. 
Hence, by shortcutting any nonempty tour $T$, we can obtain a cycle $C$ with $\V\lrA{C}=\V\lrA{T}$ and $c\lrA{C}\leq c\lrA{T}$.}

\textls[-1]{In the \ac{WBSCP}, we are given a set of $M\le N$ identical sensors with a common maximum speed $s>0$.
A vertex $v\in\V$ is associated with a required sweep period $t_v>0$, and is said to be \emph{sweep-covered} if it is visited by at least one sensor during every sliding time window of length $t_v$.
Following previous studies~\cite{LZD:24:Optim_Lett,WSZ:25:IEEE_J_NET}, we assume that all vertices share a common sweep period $t$, namely, $t_v=t$ for every $v\in\V$.
Then, during any sweep period, each sensor can travel at most}
\begin{equation}
\Delta=st.
\end{equation}

A \emph{schedule} specifies a trajectory for each sensor over time.
The schedule is \emph{feasible} if every sensor moves at speed at most $s$.
Moreover, sensors are not associated with predetermined depots and may start from arbitrary vertices.
Our algorithms construct \emph{periodic schedules}, in which each sensor repeatedly executes its assigned tour.
For any tour $T$, let
\begin{equation}
\delta\lrA{T}=
\begin{cases}
0,& \text{if $\V\lrA{T}=\emptyset$},\\
\max\lrC{1,\left\lceil c\lrA{T}/\Delta\right\rceil},&\text{otherwise},
\end{cases}
\end{equation}
where $\left\lceil x\right\rceil$ denotes the smallest integer not less than $x$.

\begin{property}[Tour Coverage] \label{prop:cycle}
For any tour $T$, the vertices in $\V\lrA{T}$ can be sweep-covered using $\delta\lrA{T}$ mobile sensors.
\end{property}

For a tour of positive cost, this property follows by uniformly distributing $\delta\lrA{T}$ sensors along the tour $T$ and letting them move at speed $s$ in the same direction.
An empty tour requires no sensor. For a nonempty tour of zero cost, one sensor remains at its only vertex.

\begin{definition}[\ac{WBSCP}]
Given a metric graph $\G=\lrA{\V,\E}$, a nonnegative weight function $w:\V\rightarrow\mathbb{R}_{\ge0}$, a fleet of $M$ identical mobile sensors moving at speed at most $s$, and a common sweep period $t$, the \ac{WBSCP} asks for a feasible schedule using at most $M$ mobile sensors so as to maximize the total weight of the sweep-covered vertices.
\end{definition}

The \ac{BSCP} is the special case of the \ac{WBSCP} in which $w\lrA{v}=1$ for every vertex $v\in\V$.

Let $\Sigma^*$ denote an optimal solution to the \ac{WBSCP}, and define $\OPT \coloneqq w\lrA{\V^*}$, where $\V^*\subseteq\V$ denotes the set of vertices sweep-covered by $\Sigma^*$.
Finally, we define the bifactor approximation used in the paper.

\begin{definition}[Bifactor Approximation Algorithm]
For any $\alpha\geq 1>\beta>0$, an algorithm is called a bifactor $\lrA{\alpha,\beta}$-approximation algorithm for the \ac{WBSCP} if, for every instance with a budget of $M$ mobile sensors, it computes in polynomial time a feasible solution using at most $\alpha M$ mobile sensors and whose total covered weight is at least $\beta\OPT$.
\end{definition}

\section{Constant-Factor Approximation via\\ Route Transformation}\label{sec:3}
In this section, we propose a constant-factor approximation algorithm for the \ac{WBSCP}.
The algorithm uses any bifactor $\lrA{\alpha,\beta}$-approximation algorithm for the \ac{WBSCP} as a black-box subroutine and transforms its solution into a feasible periodic schedule using at most $M$ sensors.
The main idea is to convert the possibly complicated collaborative trajectories generated by the bifactor approximation algorithm into a set of short, independent cycles.
Since a cycle of cost at most $\Delta$ can be sweep-covered by a single sensor, selecting a proper subset of these cycles allows us to satisfy the original sensor budget.
The approximation guarantee is obtained by showing that the total weight preserved by the selected cycles is a constant fraction of the weight covered by the original bifactor solution.

\subsection{$\beta/\lra{2\alpha}$-Approximation Algorithm}

Suppose that we are given a bifactor $\lrA{\alpha,\beta}$-approximation oracle for the \ac{WBSCP}.
By definition, it returns a feasible schedule using at most $\alpha M$ sensors and sweep-covering vertices of total weight at least $\beta\OPT$.
We first extract the trajectories generated by the obtained schedule.
Specifically, we consider the trajectory of each sensor during a time window $\left[0,t\right]$ of duration $t$.
Since each sensor travels at speed at most $s$, the sequence of vertices it visits induces a walk in $\G$ of cost at most $\Delta=st$.
Let $\W$ denote the set of walks corresponding to all deployed sensors, where $\Size{\W}\leq\alpha M$.
Although these walks may contain repeated vertices and overlap with each other, shortcutting can remove such repetitions without increasing the travel cost due to the triangle inequality.

To obtain mutually vertex-disjoint paths, we process the walks in an arbitrary order.
For each walk, we apply shortcutting to iteratively remove any repeated vertices within the walk itself, as well as any vertices already covered by previously processed paths.
If all vertices of the walk have already been covered, the walk is simply discarded.
Since every vertex whose occurrence is removed remains in the current walk or in a previously constructed path, the union of covered vertices over all paths remains unchanged.
Let $\P_{\mathrm{init}}$ be the resulting set of nonempty paths.
By construction, all paths in $\P_{\mathrm{init}}$ are pairwise vertex-disjoint, and the cost of each path remains at most $\Delta$.

Next, we transform each path into a set of short paths.
For any path $P\in\P_{\mathrm{init}}$, we greedily partition its vertex sequence into a set of vertex-disjoint subpaths such that the cost of each subpath is at most $\Delta/2$.
At each step, we take the longest prefix of the remaining vertex sequence whose cost is at most $\Delta/2$.
This bound is selected to ensure that each resulting subpath can be completed into a cycle of cost at most $\Delta$.
More precisely, for every subpath $P'$, we add an edge between its two endpoints to obtain a cycle.
By the triangle inequality, the cost of the completed cycle satisfies
\begin{equation}
c\lrA{C'}\leq 2c\lrA{P'}\leq \Delta.
\end{equation}

Let $\C$ denote the set of all cycles obtained in this way.
The above transformation decomposes the trajectories of the bifactor solution into a set of vertex-disjoint short cycles.
By Property~\ref{prop:cycle}, each cycle in $\C$ can be sweep-covered by one sensor.
Hence, it remains to select at most $M$ cycles from $\C$ while maximizing the total covered weight.
Since all cycles in $\C$ are vertex-disjoint, the weight contribution of each cycle can be evaluated independently.
Hence, we compute the weights of all cycles in $\C$, sort them in descending order, and greedily select at most $M$ cycles with the largest weights.
The selected cycles form a feasible solution to the \ac{WBSCP}.
The detailed procedure is summarized in Algorithm~\ref{alg1}.
We show in Theorem~\ref{thm1} that this procedure is a $\beta/\lra{2\alpha}$-approximation algorithm.

\begin{algorithm}[!t]
\caption{Constant-Factor Approximation for the \ac{WBSCP}}
\label{alg1}
\begin{algorithmic}[1]
\renewcommand{\algorithmicrequire}{\textbf{Input:}}
\renewcommand{\algorithmicensure}{\textbf{Output:}}
\REQUIRE The instance $\G=\lrA{\V,\E}$, fleet size $M$, distance budget $\Delta$, and a bifactor $\lrA{\alpha,\beta}$-approximation oracle for the \ac{WBSCP}.
\ENSURE A periodic schedule using at most $M$ mobile sensors.

\STATE Invoke the \ac{WBSCP} oracle to obtain a feasible schedule.\label{alg1s1}

\STATE Extract a set $\W$ of walks induced by the trajectories of all sensors during the time window $\left[0,t\right]$.\label{alg1s2}
\STATE Initialize $\P_{\mathrm{init}}\leftarrow\emptyset$ and $\C\leftarrow\emptyset$.\label{alg1s3}
\FOR{each walk $W\in\W$ in an arbitrary order}\label{alg1s4}
    \STATE Apply shortcutting to remove repeated vertices within $W$ and vertices already contained in paths in $\P_{\mathrm{init}}$.\label{alg1s15}
    \IF{the resulting path $P$ is nonempty}\label{alg1s6}
        \STATE \textls[-12]{Add $P$ to $\P_{\mathrm{init}}$, and greedily partition $P$ into vertex-disjoint subpaths $\P_{\mathrm{sub}}$ with $c\lrA{P'}\leq\Delta/2$ for each $P'\in\P_{\mathrm{sub}}$.} \label{alg1s7}
        \FOR{each subpath $P'\in\P_{\mathrm{sub}}$}\label{alg1s8}
            \STATE Complete $P'$ into a cycle $C'$, and add $C'$ to $\C$.\label{alg1s9}
        \ENDFOR
    \ENDIF
\ENDFOR
\STATE Sort $\C$ in nonincreasing order of their weights, and let $\C_{\mathrm{final}}$ be the first $\min\lrC{M,\Size{\C}}$ cycles.\label{alg1s13}
\RETURN The schedule obtained by assigning one sensor to sweep-cover each cycle in $\C_{\mathrm{final}}$.\label{alg1s14}
\end{algorithmic}
\end{algorithm}

\subsection{Approximation and Complexity Analysis}

We analyze Algorithm~\ref{alg1} in three steps. 
First, we consider the structural properties of the sequential shortcutting procedure. We then analyze the cycle decomposition step. Finally, we establish the approximation guarantee.

\begin{lemma}[Sequential Shortcutting] \label{lem:shortcut}
Let $\W$ and $\P_{\mathrm{init}}$ be the sets of walks and paths produced by Algorithm~\ref{alg1} before and after the sequential shortcutting procedure, respectively. Then the following properties hold.
\begin{enumerate} 
    \item The paths in $\P_{\mathrm{init}}$ are pairwise vertex-disjoint.
    \item $\V\lrA{\P_{\mathrm{init}}}=\V\lrA{\W}$.
    \item Every path $P\in\P_{\mathrm{init}}$ satisfies $c\lrA{P}\le\Delta$.
\end{enumerate}
\end{lemma}

\begin{IEEEproof}
The walks in $\W$ are processed sequentially by lines~\ref{alg1s4}--\ref{alg1s9} of Algorithm~\ref{alg1}. 
Whenever a walk is considered, every vertex already contained in a previously generated path is removed by the shortcutting operation in line~\ref{alg1s15}. 
Hence, no vertex can appear in two different paths in $\P_{\mathrm{init}}$, proving 1).
A vertex whose occurrence is removed remains in the current walk or in a previously generated path. 
Hence, shortcutting does not change the union of covered vertices, so $\V\lrA{\P_{\mathrm{init}}}=\V\lrA{\W}$, proving 2).
Finally, shortcutting never increases the cost. 
Since every extracted walk has cost at most $\Delta$, every resulting path also satisfies $c\lrA{P}\le\Delta$, proving 3).
\end{IEEEproof}

\begin{lemma}[Cycle Decomposition]\label{lem:decomposition}
Let $\C$ denote the set of cycles generated from $\P_{\mathrm{init}}$ in Algorithm~\ref{alg1}. Then the following properties hold.
\begin{enumerate}
    \item Every cycle $C'\in\C$ satisfies $c\lrA{C'}\le\Delta$.
    \item The cycles in $\C$ are pairwise vertex-disjoint.
    \item $\V\lrA{\C}=\V\lrA{\P_{\mathrm{init}}}$.
    \item $\Size{\C}\le2\alpha M$.
\end{enumerate}
Consequently, $w\lrA{\C}=w\lrA{\P_{\mathrm{init}}}=w\lrA{\W}$.
\end{lemma}

\begin{IEEEproof}
By line~\ref{alg1s7} of Algorithm~\ref{alg1}, each path in $\P_{\mathrm{init}}$ is partitioned into subpaths, each having cost at most $\Delta/2$. 
When we complete a subpath $P'$ into a cycle $C'$, we add one edge whose cost is at most $c\lrA{P'}$ by the triangle inequality. 
Hence, every cycle $C'\in\C$ satisfies $c\lrA{C'}\le 2c\lrA{P'}\le\Delta$, proving 1).
The paths in $\P_{\mathrm{init}}$ are pairwise vertex-disjoint by Lemma~\ref{lem:shortcut}. 
Since each path is partitioned into vertex-disjoint subpaths, the cycles in $\C$ are pairwise vertex-disjoint, proving 2).
The partition and completion steps do not change the vertex set of a path. Hence, $\V\lrA{\C}=\V\lrA{\P_{\mathrm{init}}}$, proving 3).
Regarding the cardinality bound 4), recall that the $\lrA{\alpha,\beta}$-approximation oracle uses at most $\alpha M$ mobile sensors, so $\size{\P_{\mathrm{init}}}\le\size{\W}\le\alpha M$. 
Since every path in $\P_{\mathrm{init}}$ satisfies $c\lrA{P}\le\Delta$ by Lemma~\ref{lem:shortcut}-3), the greedy partition produces at most two subpaths.
Otherwise, the first two subpaths, each together with the edge connecting it to the next subpath, would each have cost greater than $\Delta/2$.
These two parts are edge-disjoint, contradicting $c\lrA{P}\le\Delta$.
Hence, the total number of vertex-disjoint cycles generated is at most $\size{\C} \le 2\size{\P_{\mathrm{init}}} \le 2\alpha M$, proving 4).
Finally, $w\lrA{\C}=w\lrA{\W}$ follows immediately from property 3) and Lemma~\ref{lem:shortcut}-2).
\end{IEEEproof}

The next lemma analyzes the cycle selection step.

\begin{lemma}[Greedy Cycle Selection]\label{lem:selection}
The cycle set $\C_{\mathrm{final}}$ returned by Algorithm~\ref{alg1} satisfies $w\lrA{\C_{\mathrm{final}}} \ge 1/\lra{2\alpha} \, w\lrA{\C}$.
\end{lemma}

\begin{IEEEproof}
The claim is immediate if $\C=\emptyset$, so assume $\C\ne\emptyset$.
By Lemma~\ref{lem:decomposition}-2), the cycles in $\C$ are pairwise vertex-disjoint. Hence, their weights are additive. 
In line~\ref{alg1s13}, Algorithm~\ref{alg1} sorts them in nonincreasing order of their weights and selects the first $\min\lrC{M,\Size{\C}}$ cycles. 
Since $\Size{\C}\le2\alpha M$ by Lemma~\ref{lem:decomposition}-4), by an averaging argument, we obtain
\begin{equation}
w\lrA{\C_{\mathrm{final}}} \ge\frac{\min\lrC{M,\Size{\C}}}{\size{\C}}w\lrA{\C} \ge \frac{1}{2\alpha}w\lrA{\C},
\end{equation}
which completes the proof.
\end{IEEEproof}

We are now ready to establish the approximation guarantee.

\begin{theorem}[Route-Transformation Guarantee]\label{thm1}
Given any bifactor $\lrA{\alpha,\beta}$-approximation algorithm for the \ac{WBSCP}, Algorithm~\ref{alg1} is a polynomial-time $\beta/\lra{2\alpha}$-approximation algorithm for the \ac{WBSCP}. In particular, the \ac{WBSCP} admits a polynomial-time $1/16$-approximation algorithm.
\end{theorem}

\begin{IEEEproof}
By definition, the oracle returns a feasible schedule covering vertices of total weight at least $\beta\OPT$. 
By Lemmas~\ref{lem:shortcut} and \ref{lem:decomposition}, we obtain 
\begin{equation}
w\lrA{\C}=w\lrA{\W}\ge\beta\OPT.
\end{equation}
Then, by Lemma~\ref{lem:selection}, we obtain 
\begin{equation}
w\lrA{\C_{\mathrm{final}}} \ge \frac{1}{2\alpha}w\lrA{\C} \ge \frac{\beta}{2\alpha}\OPT.
\end{equation}
By Lemma~\ref{lem:decomposition}-1), every selected cycle has cost at most $\Delta$. Hence, one mobile sensor suffices to sweep-cover each selected cycle. 
Since at most $M$ cycles are selected, the returned periodic schedule is feasible and uses at most $M$ sensors.

Let $t_{\mathrm{oracle}}$ be the running time of the $\lrA{\alpha,\beta}$-approximation oracle, and let $L$ be the total number of vertex occurrences in the trajectories returned by the oracle.
The post-processing operations in lines~\ref{alg1s2}--\ref{alg1s13} of Algorithm~\ref{alg1} take $O\lrA{L+N\log N}$ time, where the $O\lrA{L}$ term is for trajectory extraction, shortcutting, path partitioning, and cycle generation, and the $O\lrA{N\log N}$ term is for sorting the generated cycles.
Note that the oracle runs in polynomial time, and the total number $L$ of vertex occurrences in its output satisfies $L=O\lrA{t_{\mathrm{oracle}}}$.
Therefore, the overall running time of Algorithm~\ref{alg1} is $O\lrA{t_{\mathrm{oracle}}+N\log N}$.

Thus, applying the bifactor $\lrA{4,1/2}$-approximation algorithm for the \ac{WBSCP}~\cite{LZD:24:Optim_Lett} yields the $1/16$-approximation.
\end{IEEEproof}



\section{Improved Approximation via Greedy\\ Maximum Coverage} \label{sec:4}

In this section, we improve the approximation ratio for the \ac{WBSCP} by proposing a $\lra{1-e^{-1/2}-\varepsilon}$-approximation algorithm for any fixed $\varepsilon>0$.
The algorithm uses a minimum-excess path subroutine and constructs a feasible solution using at most $M$ sensors.
Instead of exploiting collaborative sweep-covering among multiple sensors, our algorithm constructs $M$ independent budget tours, one for each sensor.
These tours are generated greedily by repeatedly invoking a short-path dominating oracle with vertex weights updated to reflect only the uncovered marginal weights.
The approximation is obtained by combining a short-window averaging argument for the optimal solution with an analysis of the greedy iterative process.

\subsection{A Short-Path Dominating Oracle}

We first introduce the routing subroutine in our algorithm. For two distinct vertices $a,b$, the \emph{excess} of an $a$--$b$ path $P$ is defined as $\operatorname{exc}\lrA{P}=c\lrA{P}-c\lrA{a,b}$. Given nonnegative vertex weights $p$ and a value $q>0$, the weighted $q$-minimum-excess path ($q$-\acs{MEP}) problem seeks an $a$--$b$ path of minimum excess such that the total weight of its visited vertices is at least $q$.
For this problem, a bifactor $\lra{1/\lrA{1+\varepsilon},2+\varepsilon}$-approximation algorithm is known for any fixed $\varepsilon>0$~\cite{FS:14:STOC}. It computes in polynomial time an $a$--$b$ path $R$ with $p\lrA{R}\ge q/\lra{1+\varepsilon}$ and
\begin{equation}\label{eq:sw_me_excess}
\operatorname{exc}\lrA{R}
\le
\lra{2+\varepsilon}
\min_{\substack{P: a\leadsto b,p\lrA{P}\ge q}}
\operatorname{exc}\lrA{P}.
\end{equation}

We next use this bifactor algorithm to construct the required short-path dominating oracle.
For a length bound $\ell\ge0$, let $\Phi_\ell\lrA{p}$ denote the maximum weight of a path of cost at most $\ell$, where empty and singleton paths are also allowed.

\begin{lemma}[Short-Path Dominating Oracle]\label{lem:sw_oracle}
For any budget $D>0$, nonnegative vertex weights $p$, and any fixed $\varepsilon>0$, one can compute in polynomial time a tour $T$ such that
\begin{equation}\label{eq:sw_oracle}
c\lrA{T}\le D,
\qquad
p\lrA{T}
\ge
\frac{1}{\lra{1+\varepsilon}^2}
\Phi_{D/\lra{2+\varepsilon}}\lrA{p}.
\end{equation}
\end{lemma}
\begin{IEEEproof}
We construct a set of candidate tours and return one with maximum weight. 
Let $p_{\max}=\max_{v\in\V}p\lrA{v}$. If $p_{\max}=0$, the empty tour satisfies the claim. Otherwise, we include every singleton tour as a candidate. 
For every ordered pair of distinct vertices $a,b$ and every $j=0,1,\ldots$, we consider the values $q=p_{\max}\lra{1+\varepsilon}^j\le p\lrA{\V}$. 
For each such value, we apply the bifactor $q$-\ac{MEP} algorithm of~\cite{FS:14:STOC}, complete the returned path, and retain the resulting tour if its cost is at most $D$.

Let $P^*$ be a path attaining $\Phi_{D/\lra{2+\varepsilon}}\lrA{p}$. If $P^*$ is a singleton, it is already among the candidates. Otherwise, let $a,b$ be its endpoints. Since every singleton path is feasible in the definition of $\Phi_{D/\lra{2+\varepsilon}}\lrA{p}$, we have $p_{\max}\le p\lrA{P^*}\le p\lrA{\V}$. Hence, some value $q$ considered above satisfies $q\le p\lrA{P^*}<\lra{1+\varepsilon}q$, and then $P^*$ is feasible for the corresponding $q$-\ac{MEP} instance.

Let $R$ denote the path returned for this instance. We complete $R$ into a tour. By \eqref{eq:sw_me_excess}, we obtain
\begin{align}
c\lrA{R}+c\lrA{a,b}
&\le \lra{2+\varepsilon}\lrA{c\lrA{P^*}-c\lrA{a,b}}+2c\lrA{a,b}\nonumber\\
&= \lra{2+\varepsilon}c\lrA{P^*}-\varepsilon c\lrA{a,b}
\le D.
\label{eq:sw_closing}
\end{align}
Thus, this tour is retained. Moreover, the bifactor guarantee and the choice of $q$ imply $p\lrA{R}\ge q/\lra{1+\varepsilon}>p\lrA{P^*}/\lra{1+\varepsilon}^2$. Hence, the maximum-weight retained candidate satisfies \eqref{eq:sw_oracle}.

Since $p(\V)\le Np_{\max}$, only polynomially many values of $q$ are considered for each ordered pair of endpoints.
Moreover, since there are at most $N^2$ such pairs, the construction runs in polynomial time for any fixed $\varepsilon>0$.
\end{IEEEproof}

\subsection{Greedy Maximum-Coverage Algorithm}

Algorithm~\ref{alg2} constructs the solution greedily over $M$ iterations.
Initially, no vertex has been covered, i.e., $\U_0=\emptyset$.
After the first $j-1$ iterations, let $\U_{j-1}$ be the set of vertices already covered by the selected tours.
To avoid repeatedly collecting weight from previously covered vertices, we define the marginal weight function
\begin{equation}
w_j(v)=
\begin{cases}
w\lrA{v}, & \text{if $v\notin \U_{j-1}$},\\
0, & \text{otherwise}.
\end{cases}
\end{equation}
Hence, each uncovered vertex preserves its weight, while each previously covered vertex contributes zero marginal weight.
The oracle in Lemma~\ref{lem:sw_oracle} is then invoked with the updated weight function $w_j$ and distance budget $\Delta$ to compute a budget tour $T_j$.
The tour is added to the current solution, and the covered vertex set is updated by $\U_j=\U_{j-1}\cup\V\lrA{T_j}$.
This procedure is repeated until $M$ tours have been generated.
After the final iteration, the algorithm returns the tour set $\T=\lrC{T_1,\ldots,T_M}$.
Since every tour produced by the oracle satisfies the budget constraint $c\lrA{T_j}\le\Delta$, assigning one sensor to each nonempty tour yields a feasible periodic schedule for the \ac{WBSCP}.

\begin{algorithm}[!t]\color{black}
\caption{Improved Approximation for the \ac{WBSCP}}
\label{alg2}
\begin{algorithmic}[1]
\renewcommand{\algorithmicrequire}{\textbf{Input:}}
\renewcommand{\algorithmicensure}{\textbf{Output:}}
\REQUIRE The instance $\G=\lrA{\V,\E}$, fleet size $M$, distance budget $\Delta$, a fixed parameter $\varepsilon>0$, and the oracle in Lemma~\ref{lem:sw_oracle}.
\ENSURE A periodic schedule using at most $M$ mobile sensors.

\STATE Initialize $\U_0\leftarrow\emptyset$ and $\T\leftarrow\emptyset$.\label{alg2s1}
\FOR{$j=1,\ldots,M$}\label{alg2s2}
    \STATE Define the marginal weight function $w_j$ by setting $w_j\lrA{v} = w\lrA{v}$ if $v \notin \U_{j-1}$, and $w_j\lrA{v} = 0$ otherwise.\label{alg2s3}

    \STATE Invoke the oracle in Lemma~\ref{lem:sw_oracle} on $\G$ with budget $\Delta$ and weight function $w_j$ to obtain a tour $T_j$.\label{alg2s4}
    \STATE Add $T_j$ to $\T$, and update $\U_j \leftarrow \U_{j-1} \cup \V\lrA{T_j}$.\label{alg2s5}
\ENDFOR
\RETURN The schedule obtained by assigning one sensor to sweep-cover each nonempty tour in $\T$.\label{alg2s7}
\end{algorithmic}
\end{algorithm}

The detailed procedure is summarized in Algorithm~\ref{alg2}.

\subsection{Approximation and Complexity Analysis}

We analyze Algorithm~\ref{alg2} in three steps.
First, we relate the weight covered by a feasible schedule to its trajectories in a short time window.
We then bound the marginal gain in each greedy iteration.
Finally, we establish the approximation ratio.

\begin{lemma}[Short-Window Coverage]\label{lem:sw_window}
Let $\Sigma$ be any feasible schedule with maximum speeds $s_1,\ldots,s_M$ sweep-covering $U$.
For any nonnegative vertex weights $p$ and $0<\tau\le t$, we have
\begin{equation}\label{eq:sw_group_witness}
\sum_{i=1}^M \Phi_{s_i\tau}\lrA{p}\ge\frac\tau t p\lrA{U}.
\end{equation}
\end{lemma}
\begin{IEEEproof}
Fix $v\in U$ and choose a starting time $a$ uniformly from $[0,H]$. Consider the maximal intervals in $[0,H+\tau]$ during which no sensor visits $v$, and let their lengths be $g_1,g_2,\ldots$. Since $v$ is sweep-covered, each such interval has length at most $t$.

The window $[a,a+\tau]$ avoids $v$ only if it lies entirely within one of these intervals. 
Since $a$ is chosen uniformly from $[0,H]$, 
\begin{align}\label{eq:sw_probability}
\Pr\lrB{v\text{ is visited in }[a,a+\tau]} &\geq 1 - \frac{\sum_{j}\max\lrc{g_j-\tau,0}}{H}\nonumber\\
&\ge \tau/t-O\lrA{t/H},
\end{align}
where the last inequality follows from $\max\lrC{g-\tau,0}\le\lra{1-\tau/t}g$ for $0\le g\le t$ and $\sum_jg_j\le H+\tau$.

By linearity of expectation, the expected total weight of the vertices in $U$ visited during the window is at least $\lra{\tau/t-O\lrA{t/H}}p\lrA{U}$. 
For any fixed starting time $a$, sensor $i$ travels a distance of at most $s_i\tau$ during $[a,a+\tau]$. We shortcut its visited vertices to obtain a path of cost at most $s_i\tau$, and hence the weight visited by sensor $i$ is at most $\Phi_{s_i\tau}\lrA{p}$.
Therefore, the total weight of the vertices in $U$ visited by all sensors during the window is at most $\sum_{i=1}^M \Phi_{s_i\tau}\lrA{p}$.

Combining the expected lower bound with this upper bound and letting $H\to\infty$ yields $\sum_{i=1}^M \Phi_{s_i\tau}\lrA{p} \ge \frac{\tau}{t}p\lrA{U}$.
\end{IEEEproof}

Let $G_0 = 0$ and let $G_j = w\lrA{\U_j}$ for each $j\in\lrB{M}$ denote the total weight covered by Algorithm~\ref{alg2} after the first $j$ iterations. The following lemma establishes a lower bound on the marginal gain achieved in each iteration.

\begin{lemma}[Greedy Progress] \label{lem:greedy_step}
For each iteration $j\in\lrB{M}$, we have $G_j-G_{j-1}\ge\frac{\beta}{M}\lrA{\OPT-G_{j-1}}$, where $\beta=\frac{1}{\lra{1+\varepsilon}^2\lra{2+\varepsilon}}$.
\end{lemma}
\begin{IEEEproof}
Recall that $\Sigma^*$ denotes an optimal solution to the \ac{WBSCP}.
For the marginal weight function $w_j$, we have
\begin{equation}
w_j\lrA{\V^*}=\OPT-w\lrA{\V^*\cap\U_{j-1}}
\ge\OPT-G_{j-1}.
\end{equation}
Apply Lemma~\ref{lem:sw_window} to $\Sigma^*$ with $\tau=t/\lra{2+\varepsilon}$. Since all sensors have maximum speed $s$ and $\Delta=st$, we have $s\tau=\Delta/\lra{2+\varepsilon}$. Hence,
\begin{equation}
\Phi_{\Delta/\lra{2+\varepsilon}}\lrA{w_j}\ge\frac{\OPT-G_{j-1}}{\lra{2+\varepsilon}M}.
\end{equation}
By Lemma~\ref{lem:sw_oracle}, the tour computed in line~\ref{alg2s4} satisfies
\begin{equation}
w_j\lrA{T_j}\ge\lra{1+\varepsilon}^{-2}\Phi_{\Delta/\lra{2+\varepsilon}}\lrA{w_j}\ge\frac{\beta}{M}\lrA{\OPT-G_{j-1}}.
\end{equation}
Finally, since $G_j-G_{j-1}=w_j\lrA{T_j}$, the lemma follows.
\end{IEEEproof}

We are now ready to establish the approximation guarantee.

\begin{theorem}[Greedy Coverage Guarantee] \label{thm3}
For any fixed $\varepsilon>0$, the \ac{WBSCP} admits a polynomial-time $\lra{1-e^{-1/2}-\varepsilon}$-approximation algorithm.
\end{theorem}
\begin{IEEEproof}
We first analyze Algorithm~\ref{alg2}.
By Lemma~\ref{lem:greedy_step}, for every iteration $j\in\lrB{M}$, we have
\begin{align}
\OPT-G_j
&\le\OPT-G_{j-1}-\frac\beta M\lrA{\OPT-G_{j-1}}\nonumber\\
&=\lrA{1-\beta/M}\lrA{\OPT-G_{j-1}}.
\end{align}
Applying this inequality recursively for all iterations in lines~\ref{alg2s2}--\ref{alg2s5} of Algorithm~\ref{alg2} and using $G_0=0$, we obtain
\begin{equation}\label{eq:opt}
\OPT-G_M\le\lrA{1-\beta/M}^M\OPT\le e^{-\beta}\OPT,
\end{equation}
where the last inequality follows from $\lrA{1-x}^m\le e^{-mx}$ for $0\le x\le1$. Therefore, by \eqref{eq:opt}, we have
\begin{equation}\label{eqalg2}
G_M\ge\lrA{1-e^{-\beta}}\OPT.
\end{equation}

By line~\ref{alg2s4}, every computed tour has cost at most $\Delta$.
Hence, assigning one sensor to each nonempty tour returned in line~\ref{alg2s7} yields a feasible periodic schedule using at most $M$ sensors, and this schedule achieves an approximation ratio of $1-e^{-\beta}$.

For the running time, let $t'{\!}_{\mathrm{oracle}}$ denote the running time of Lemma~\ref{lem:sw_oracle}.
Since the oracle is invoked once in each of the $M$ iterations, the total oracle time is $M t'{\!}_{\mathrm{oracle}}$.
The remaining operations in lines~\ref{alg2s1}--\ref{alg2s5}, including updating the marginal weight function and maintaining the covered vertex set, take $O\lrA{MN}$ time in total.
Hence, the overall running time of Algorithm~\ref{alg2} is $O\lrA{M t'{\!}_{\mathrm{oracle}}+MN}$.

Finally, since $\beta\to1/2$ as the accuracy parameter of Algorithm~\ref{alg2} tends to zero, for any fixed $\varepsilon>0$ this parameter can be chosen sufficiently small so that \eqref{eqalg2} yields an approximation ratio of at least $1-e^{-1/2}-\varepsilon$.
\end{IEEEproof}

\section{Extension to Heterogeneous Sensor Speeds}\label{sec:5}

The algorithms in the previous sections assume that all sensors have the same maximum speed and therefore share the same travel budget within one sweep period. In practice, sensors may have heterogeneous capabilities due to differences in hardware, battery conditions, or mobility constraints~\cite{LS:21:Sensors}. We now extend our approach to this setting.

The main difficulty is that each sensor has its own travel budget. Thus, a tour feasible for one sensor may be infeasible for another, and the greedy construction in Algorithm~\ref{alg2} can no longer optimize over a common set of feasible tours. To handle these sensor-dependent feasible regions, we view each sensor as one group and the coverage sets of its feasible tours as the sets in that group. We then combine the short-window bound with a group-coverage lemma to obtain the same $\lra{1-e^{-1/2}-\varepsilon}$-approximation ratio as in the homogeneous setting.

\subsection{Problem Formulation and Group Coverage}

In the heterogeneous setting, the fleet consists of $M$ sensors with individual maximum speeds $s_1,s_2,\ldots,s_M>0$. Given a common sweep period $t$, sensor $i$ has distance budget $\Delta_i=s_it$.

\begin{definition}[\ac{HWBSCP}]
Given a metric graph $\G=\lrA{\V,\E}$, a nonnegative vertex weight function $w:\V\rightarrow\R_{\ge0}$, a fleet of $M$ mobile sensors with maximum speeds $s_1,s_2,\ldots,s_M>0$, and a common sweep period $t$, the \ac{HWBSCP} asks for a feasible schedule using these sensors so as to maximize the total weight of the sweep-covered vertices.
\end{definition}

Let $\OPT_{\rm h}$ denote the optimal objective value of the \ac{HWBSCP}. 
To capture the sensor-dependent sets of feasible tours, we use the \ac{GBMCP}~\cite{CA:04:APPROX}.

\begin{definition}[\ac{GBMCP}]
\textls[-1]{Given a ground element set $\I$, a nonnegative weight function $w:\I\rightarrow\R_{\ge0}$, a collection $\S$ (possibly containing copies) of subsets of $\I$, and a partition of $\S$ into groups $\A_1,\ldots,\A_k$, the \ac{GBMCP} asks for a subset $\S'\subseteq\S$ maximizing $\sum_{e\in \bigcup_{S\in\S'}S} w\lrA{e}$ subject to selecting at most one subset from each group, namely, $\size{\S'\cap\A_i}\le1$ for $i\in\lrB{k}$.} 
\end{definition}

\textls[-1]{Since the number of subsets in a group can be superpolynomial in $\size{\I}$, the groups are generally not given explicitly.
Instead, we assume access to an approximation oracle that, given a group and the current element weights, returns an approximate solution for that group.}

For the analysis, we introduce an additional comparison family $\B_i$ for each group $i$. 

\begin{lemma}[Group Coverage]\label{lem:dominating_coverage}
Let $\I$ be a ground set of elements with nonnegative weights $w$.
For each $i\in\lrB{K}$, let $\A_i\subseteq2^{\I}$ be a nonempty group of feasible sets and let $\B_i\subseteq2^{\I}$ be its nonempty comparison family.
Let $x\in[0,1]^{\size{\I}}$, $0\le\lambda\le1$, and $0<\gamma\le1$ be fixed. For any nonnegative weight function $p$, suppose that the oracle computes in polynomial time a set $S_i\in\A_i$ for each $i\in\lrB{K}$ such that
\begin{equation}\label{eq:dominating_oracle}
p\lrA{S_i}\ge\gamma\max_{S\in\B_i}p\lrA{S},
\end{equation}
and that
\begin{equation}\label{eq:dominating_witness}
\sum_{i=1}^K\max_{S\in\B_i}p\lrA{S}
\ge\lambda\sum_{v\in\I}p\lrA{v}x_v.
\end{equation}
Then, for any fixed $\varepsilon>0$, there exists a polynomial-time algorithm that selects at most one set from each $\A_i$ whose union has weight at least
\begin{equation}\label{eq:dominating_value}
\lrA{1-e^{-\gamma\lambda}-\varepsilon}
\sum_{v\in\I}w\lrA{v}x_v.
\end{equation}
\end{lemma}
\begin{proof}
The proof is given in Section~\ref{sec:group_coverage}.
\end{proof}

\subsection{Algorithm via Group Coverage}

Following Algorithm~\ref{alg2}, we construct at most one independent tour for each sensor.
For each sensor $i\in\lrB{M}$, let $\T_i=\lrC{T:c\lrA{T}\le\Delta_i}$ denote the set of all tours that can be executed by sensor $i$ within one sweep period.
For each tour $T\in\T_i$, we associate a coverage set $\S_T=\V\lrA{T}$.

The constructed \ac{GBMCP} instance is defined as follows.
\textls[-1]{The ground element set is the vertex set $\V$, and each vertex inherits its weight from the original \ac{HWBSCP}.
For each sensor $i$, all subsets generated by tours in $\T_i$ form one group $\A_i = \lrC{\S_T:T\in\T_i}$.
If identical subsets appear in different groups, they are treated as distinct copies associated with different sensors.
The objective of the resulting \ac{GBMCP} instance is to select at most one subset from each group while maximizing the total weight of the union of the selected subsets.
Hence, selecting at most one subset from each group corresponds exactly to assigning at most one feasible tour to each sensor.}

For a fixed constant $\varepsilon>0$, define the comparison family for sensor $i$ by
\begin{equation}\label{eq:het_inner}
\B_i=\lrC{\V\lrA{P}:c\lrA{P}\le\frac{\Delta_i}{2+\varepsilon}},
\end{equation}
where empty and singleton paths are allowed.

For any nonnegative vertex weights $p$, Lemma~\ref{lem:sw_oracle}, applied with budget $\Delta_i$, returns a feasible tour $T_i\in\T_i$ such that
\begin{equation}\label{eq:het_oracle}
p\lrA{T_i}\ge\gamma\max_{S\in\B_i}p\lrA{S},
\qquad \gamma=\lra{1+\varepsilon}^{-2}.
\end{equation}
Thus, $\V\lrA{T_i}\in\A_i$ satisfies the oracle condition of Lemma~\ref{lem:dominating_coverage}.

We apply Lemma~\ref{lem:dominating_coverage} using the oracle above. The selected coverage sets are then replaced by their stored tours and assigned to their respective sensors.

The procedure is summarized in Algorithm~\ref{alg3}.

\begin{algorithm}[!t]\color{black}
\caption{Group-Coverage Algorithm for the \ac{HWBSCP}}
\label{alg3}
\begin{algorithmic}[1]
\renewcommand{\algorithmicrequire}{\textbf{Input:}}
\renewcommand{\algorithmicensure}{\textbf{Output:}}
\REQUIRE The instance $\G=\lrA{\V,\E}$, fleet size $M$, distance budgets $\lrc{\Delta_i}_{i\in\lrB{M}}$, a constant $\varepsilon>0$, and the oracle in Lemma~\ref{lem:sw_oracle}.
\ENSURE A periodic schedule using at most $M$ sensors.

\STATE Define the groups $\A_1,\ldots,\A_M$ and comparison families $\B_1,\ldots,\B_M$ implicitly as above.\label{alg3s1}

\STATE Invoke Lemma~\ref{lem:dominating_coverage} using the oracle in \eqref{eq:het_oracle} for each group.\label{alg3s2}

\STATE Let $\T_{\rm out}$ denote the stored tours corresponding to the selected nonempty coverage sets.\label{alg3s3}

\RETURN The schedule obtained by assigning each selected tour to its corresponding sensor.\label{alg3s4}

\end{algorithmic}
\end{algorithm}

\subsection{Approximation and Complexity Analysis}

The same short-window argument used for the homogeneous problem applies here.
For any feasible heterogeneous schedule sweep-covering $U$, Lemma~\ref{lem:sw_window} with $\tau=t/\lra{2+\varepsilon}$ yields
\begin{equation}\label{eq:het_window}
\sum_{i=1}^M\max_{S\in\B_i}p\lrA{S}
=\sum_{i=1}^M\Phi_{\Delta_i/\lra{2+\varepsilon}}\lrA{p}
\ge\frac{p\lrA{U}}{2+\varepsilon}
\end{equation}
for every nonnegative $p$.
The common time window has sensor-specific length bounds $s_i\tau=\Delta_i/\lra{2+\varepsilon}$.

\begin{theorem}[Group Coverage Guarantee]\label{thm5}
For any fixed $\varepsilon>0$, the \ac{HWBSCP} admits a polynomial-time $\lra{1-e^{-1/2}-\varepsilon}$-approximation algorithm.
\end{theorem}
\begin{IEEEproof}
Apply \eqref{eq:het_window} to an optimal heterogeneous schedule and let $x$ assign one to each sweep-covered vertex and zero to every other vertex.
Equation~\eqref{eq:dominating_witness} holds with $K=M$ and $\lambda=\lra{2+\varepsilon}^{-1}$.
By \eqref{eq:het_oracle}, the oracle factor is $\gamma=\lra{1+\varepsilon}^{-2}$.
Lemma~\ref{lem:dominating_coverage} then yields
\begin{equation}\label{eq:het_factor}
w\lrA{\T_{\rm out}}\ge\lrA{1-e^{-\gamma\lambda}-\varepsilon}\OPT_{\rm h}.
\end{equation}
Each sensor receives at most one tour, and every selected tour in its group has cost at most $\Delta_i$.
Hence, we assign each selected tour to its corresponding sensor to obtain a feasible solution.

By Lemmas~\ref{lem:sw_oracle} and \ref{lem:dominating_coverage}, Algorithm~\ref{alg3} runs in polynomial time.
Finally, since $\gamma\lambda\to1/2$ as the accuracy parameter tends to zero, for any fixed $\varepsilon>0$ this parameter can be chosen sufficiently small so that \eqref{eq:het_factor} yields an approximation ratio of at least $1-e^{-1/2}-\varepsilon$.
\end{IEEEproof}

We conclude by comparing our framework with the existing result in~\cite{LS:21:Sensors}. 
For line metrics, a randomized $\lra{1-e^{-1}}/2$-approximation algorithm for the \acs{HWBSCP} based on randomized \ac{LP} rounding is first given in~\cite{LS:21:Sensors}. 
A deterministic algorithm with the same guarantee is subsequently obtained through a dedicated derandomization procedure.
Our group-coverage method admits a stronger guarantee on line and cycle metrics, where distances are the shortest-path distances on a given undirected weighted path or cycle, respectively.

For these metrics, each sensor's group $\A_i$ can be restricted to a polynomial-size dominating subfamily without affecting the optimum of the \ac{GBMCP}.
For a line metric, we enumerate the intervals between all pairs of vertices at distance at most $\Delta_i/2$.
For a cycle metric, we enumerate both arcs between each pair of vertices and retain those of length at most $\Delta_i/2$, together with the full cycle when its length is at most $\Delta_i$.
Singleton sets are also included.
Therefore, a capacity-two \ac{LP} argument yields the following guarantee without an additive $\varepsilon$ loss.

\begin{theorem}[Line and Cycle Metrics] \label{cor:plus}
The \acs{HWBSCP} admits a polynomial-time approximation algorithm with a ratio of $\lra{1-e^{-1/2}} > 0.3934$ on both line and cycle metrics.
\end{theorem}
\begin{IEEEproof}
Apply the trajectory extraction and completion steps of Algorithm~\ref{alg1} to each sensor's nonempty walks over two consecutive time windows of duration $t/2$ in an optimal heterogeneous schedule.
This yields at most two tours per sensor, each of cost at most $\Delta_i$, whose union covers all optimally sweep-covered vertices.
On a line, a feasible tour is contained in a feasible interval.
On a cycle, it is contained in a feasible arc traversed in both directions, unless the full cycle itself is feasible.
Thus, these tours can be replaced by supersets in the explicitly enumerated groups without losing weight.

We solve the following capacity-two coverage \ac{LP}:
\begin{align}
\max\quad&\sum_{v\in\V}w\lrA{v}x_v\label{eq:line_cycle_lp}\\
\text{s.t.}\quad&0\le x_v\le1, &&v\in\V,\nonumber\\
&x_v\le\sum_{i=1}^M\sum_{S\in\A_i:v\in S}z_{i,S}, &&v\in\V,\nonumber\\
&\sum_{S\in\A_i}z_{i,S}\le2, &&i\in\lrB{M},\nonumber\\
&z_{i,S}\ge0.&&\nonumber
\end{align}
By construction, the optimum of this \ac{LP} is at least $\OPT_{\rm h}$.
Let $(x,z)$ be an optimal solution to this \ac{LP}.

We apply the following \ac{LP} rounding.
Set $y_{i,S}=z_{i,S}/2$ and, independently for each sensor $i$, select one set $S\in\A_i$ with probability $y_{i,S}$, selecting no set with the remaining probability.
For $q_{i,v}=\sum_{S\in\A_i:v\in S}y_{i,S}$, we have $\sum_iq_{i,v}\ge x_v/2$ and
\begin{align}
\Pr\lrB{v\text{ is covered}}
&=1-\prod_i\lra{1-q_{i,v}}\nonumber\\
&\ge1-e^{-x_v/2}\ge\lra{1-e^{-1/2}}x_v.\nonumber
\end{align}
Therefore, by linearity of expectation, the expected total weight of the covered vertices is at least
\begin{align*}
\sum_{v\in\V}w\lrA{v}\Pr\lrB{v\text{ is covered}}&\ge\lra{1-e^{-1/2}}\sum_{v\in\V}w\lrA{v}x_v\\
&\ge\lra{1-e^{-1/2}}\OPT_{\rm h}.
\end{align*}
Finally, the standard method of conditional expectation~\cite{LS:21:Sensors} derandomizes the rounding in polynomial time while preserving the same guarantee. (Also see the proof of Lemma~\ref{lem:dominating_coverage}.)
\end{IEEEproof}

\section{A General Group-Coverage Framework}\label{sec:group_coverage}

In this section, we prove Lemma~\ref{lem:dominating_coverage}.
The algorithm first constructs candidate sets using the oracle in \eqref{eq:dominating_oracle}, and then selects one candidate from each group by conditional expectation.

\subsection{Construction of Candidate Sets}

We may assume that $0<\varepsilon<1$. 
Set $h=1/\ceil{1/\varepsilon}$, so $0<h\le\varepsilon$.
Initially, set $p=w$.
For each $r=0,h,\ldots,1-h$, process the groups $i=1,\ldots,K$ in order.
For group $i$, invoke the oracle with the current weights $p$, store the returned set $S_{i,r}\in\A_i$ as a candidate, and immediately update
\begin{equation}\label{eq:dc_update}
p\lrA{v}\leftarrow\lra{1-h}p\lrA{v},
\qquad v\in S_{i,r}.
\end{equation}
The weights of all other elements remain unchanged.

Assign probability $h$ to each occurrence of a stored candidate within its group.
For $S\in\A_i$, its total probability is $y_{i,S}=h\,\size{\lrC{r:S_{i,r}=S}}$.
Since each group has $1/h$ candidate occurrences, we have $\sum_{S\in\A_i}y_{i,S}=1$.
Thus, the stored candidates define a probability distribution for each group.

\subsection{Approximation Analysis}

Fix $x,\lambda$ satisfying \eqref{eq:dominating_witness}, and let $W_x=\sum_{v\in\I}w\lrA{v}x_v$ and $\beta=\gamma\lambda\in[0,1]$.
Let $p^{\lrA{r}}$ denote the weights at the beginning of iteration $r$, and let $p^{\lrA{1}}$ denote the final weights.
Define the total decrease in weights by $H_r=\sum_{v\in\I}\lrA{w\lrA{v}-p^{\lrA{r}}\lrA{v}}$, so $H_0=0$.
Write $p^{\lrA{r,i}}$ for the weights immediately before processing group $i$ in iteration $r$.
Since the weights only decrease, $p^{\lrA{r,i}}\lrA{v}\ge p^{\lrA{r+h}}\lrA{v}$ for every $v\in\I$.
By \eqref{eq:dc_update}, the oracle guarantee, and \eqref{eq:dominating_witness}, we obtain
\begin{align}
H_{r+h}-H_r
&=h\sum_{i=1}^K p^{\lrA{r,i}}\lrA{S_{i,r}}\nonumber\\
&\ge h\gamma\sum_{i=1}^K\max_{S\in\B_i}p^{\lrA{r,i}}\lrA{S}\nonumber\\
&\ge h\gamma\sum_{i=1}^K\max_{S\in\B_i}p^{\lrA{r+h}}\lrA{S}\nonumber\\
&\ge h\beta\sum_{v\in\I}p^{\lrA{r+h}}\lrA{v}x_v\nonumber\\
&\ge h\beta\lrA{W_x-H_{r+h}},\label{eq:dc_progress}
\end{align}
where the last inequality follows from $0\le x_v\le1$.
By rearranging this inequality, we obtain
\begin{equation}\label{eq:dc_recurrence}
W_x-H_{r+h}\le\lra{1+\beta h}^{-1}\lrA{W_x-H_r}.
\end{equation}
By applying \eqref{eq:dc_recurrence} for $r=0,h,\ldots,1-h$, we obtain
\begin{align}\label{eq:dc_value}
H_1&\ge\lrB{1-\lra{1+\beta h}^{-1/h}}W_x\nonumber\\
&\ge\lrA{1-e^{-\beta+\beta^2h/2}}W_x\nonumber\\
&\ge\lrA{1-e^{-\beta}-h/2}W_x\nonumber\\
&\ge\lrA{1-e^{-\gamma\lambda}-\varepsilon}W_x.
\end{align}
Here, the second inequality uses $\log\lra{1+u}\ge u-u^2/2$ for $u\ge0$.
The third follows from $\beta^2h/2\le h/2$ and the upper bound of $1$ on the derivative of $1-e^{-u}$ for $u\ge0$, since $0\le\beta-\beta^2h/2\le\beta$.

\subsection{Group Selection and Complexity Analysis}

Independently for each group $i$, select a stored candidate according to the distribution $y_{i,S}$.
For each element $v$, let $q_{i,v}=\sum_{S\in\A_i:v\in S}y_{i,S}$ denote the probability that group $i$ covers $v$.
Since every candidate occurrence has probability $h$, we have $1-q_{i,v}\le\prod_{r:v\in S_{i,r}}\lra{1-h}$, using $1-mh\le\lra{1-h}^m$ for integers $m\ge0$.
Consequently,
\begin{align}\label{eq:dc_final_weight}
w\lrA{v}\Pr\lrB{v\text{ is not covered}}
&=w\lrA{v}\prod_{i=1}^K\lra{1-q_{i,v}}\nonumber\\
&\le w\lrA{v}\prod_{i,r:v\in S_{i,r}}\lra{1-h}\nonumber\\
&=p^{\lrA{1}}\lrA{v}.
\end{align}
By summing \eqref{eq:dc_final_weight} over all elements, we conclude that the expected covered weight is at least $H_1$.

To derandomize the selection, we use conditional expectation~\cite{CM:18:ALG} to fix the group choices in order.
If earlier groups have covered $U$, choose the stored candidate $S\in\A_i$ maximizing
\begin{equation}\label{eq:dc_conditional}
\sum_{v\in S\setminus U}w\lrA{v}\prod_{j>i}\lra{1-q_{j,v}}.
\end{equation}
This is the part of the conditional expected coverage that depends on $S$.
Its maximum is at least its average under $y_{i,S}$, so this choice does not decrease the conditional expected coverage.
After all groups have been fixed, the covered weight is at least $H_1$.
Together with \eqref{eq:dc_value}, this proves \eqref{eq:dominating_value}.

For fixed $\varepsilon$, the construction stores only polynomially many candidates, with at most $1/h$ candidates per group.
Since the oracle runs in polynomial time, both candidate construction and group selection run in polynomial time.

\section{Conclusion}\label{sec:7}

In this paper, we have resolved an open question on the approximability of the \ac{WBSCP} by establishing constant-factor approximation algorithms for the problem and its variants.
Specifically, we have introduced a structural transformation from bifactor approximations, yielding a $1/16$-approximation algorithm.
We then developed a greedy maximum-coverage framework that improves the approximation to $1-e^{-1/2}-\varepsilon$ for any fixed $\varepsilon>0$ for both the \ac{WBSCP} and the \ac{BSCP}. We further extended the framework to heterogeneous mobile sensors with different travel speeds by using a group-coverage formulation, obtaining the same approximation ratio.
In the future, it will be interesting to further improve the approximation ratios for these problems. 
As a first step, one may investigate whether the approximation ratio for the \ac{HWBSCP} can be improved beyond $1-e^{-1/2}$ on line metrics.
Another interesting direction is to study more practical models, such as dynamic settings with time-varying target weights and mobile sensors subject to additional energy or depot constraints.

\section*{Acknowledgment}
In an initial version, using the $1/2$-approximation algorithm for the \acl{BTCP}~\cite{ADADD:20:MOR} as an oracle for \ac{GBMCP}, we obtained a $\lra{1-e^{-1/4}}$-approximation algorithm for both the \ac{WBSCP} and the \ac{HWBSCP}. In that version, we also considered a variant of the \ac{WBSCP}, where the trajectories of all mobile sensors are required to form a set of vertex-disjoint cycles. For this variant, by exactly (respectively., approximately) enumerating the cycle configurations, i.e., the number of cycles assigned to each number of mobile sensors, and using the same approach, we obtained a subexponential-time $\lra{1-e^{-1/2}}$-approximation algorithm (respectively., a quasipolynomial-time bifactor $\lra{1+\varepsilon,1-e^{-1/2}}$-approximation algorithm). 

Recently, discussions with GPT-6 Pro suggested the oracle in Lemma~\ref{lem:sw_oracle}, which enabled us to improve the approximation ratio to $1-e^{-1/2}-\varepsilon$ for both the \ac{WBSCP} and the \ac{HWBSCP}. Consequently, we have removed the previous results on the vertex-disjoint cycles variant from this version.

\balance


\end{document}